\documentclass[twocolumn,showpacs,preprintnumbers,amsmath,amssymb]{revtex4}
\usepackage{graphicx}
\usepackage{dcolumn}
\usepackage{bm}

\begin{document}

\title{Cooling many particles to very low temperatures}

\author{Almut Beige$^{1,2}$} \email{a.beige@imperial.ac.uk}
\author{Peter L. Knight$^1$} \email{p.knight@imperial.ac.uk}
\author{Giuseppe Vitiello$^3$} \email{vitiello@sa.infn.it}

\affiliation{$^1$Blackett Laboratory, Imperial College, Prince Consort Road, London SW7 2BW, United Kingdom} 
\affiliation{$^2$Department of Applied Mathematics and Theoretical Physics, \\ University of Cambridge, Wilberforce Road, Cambridge CB3 0WA, United Kingdom}
\affiliation{$^3$Dipartimento di Fisica ``E. R. Caianiello,'' I.N.F.N. and I.N.F.M., Universit\'a di Salerno, 84100 Salerno, Italy}

\date{\today}

\begin{abstract}
In a recent paper [Beige, Knight, and Vitiello, quant-ph/0404160], we showed that a large number $N$ of particles can be cooled very efficiently using a bichromatic interaction. The particles should be excited by red-detuned laser fields while coupling to the quantized field mode inside a resonant and leaky optical cavity. When the coupling constants are for all particles the same, a collective behavior can be generated and the cooling rate can be as large as $\sqrt{N}$ times the single-particle coupling constants. Here we study the algebraic structure of the dynamics and the origin of the collective cooling process in more detail.
\end{abstract}

\maketitle

\section{Introduction}

Collective features in the microscopic dynamics often lead to the emergence of surprising and unexpected effects in the evolution of a physical system at the macroscopic level \cite{Anderson}. The collective cooling of many two-level particles to very low temperatures is discussed here as an example of such a macroscopic manifestation of microscopic collective behavior \cite{cool}. It is shown that the collective behavior of a large number of particles can produce much higher cooling rates than they could be achieved by means of individual cooling based on the spontaneous decay of the individual particles \cite{diedrich,lewen,morigi}.

As in laser sideband cooling techniques for single two-level atoms \cite{diedrich}, we consider an experimental setup, where red-detuned laser fields increase the excitation of the particles, thereby continuously reducing the number of phonons. Afterwards, the phonon energy can be removed constantly from the system. This requires energy dissipation and yields an overall decrease of the von Neumann entropy in the setup \cite{Bartana}. One possible decay channel is spontaneous emission from the excited states of the particles. During such a photon emission, a particle returns most likely into its ground state without regaining the phonon energy lost in the excitation process. The net result is a {\it conversion} of the phonons, originally existing in the setup in the form of thermal energy, into photons escaping the system.

\begin{figure}
\begin{center}
\begin{tabular}{c}
\includegraphics[height=1.8cm]{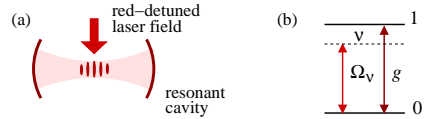}  \\[-0.2cm]
\end{tabular}
\end{center}
\caption[example]{\small (a) Experimental setup for the collective cooling of many particles. The particles should be trapped inside an optical cavity, where they arrange themselves in the antinodes of the resonator  field. (b) Level configuration of a single two-level particle for the cooling of a vibrational mode with frequency $\nu$ driven by a red-detuned laser field with Rabi frequency $\Omega_\nu$ and coupling to a resonant optical cavity with strength $g$.} \label{setup}
\end{figure}

Here we are interested in realizing much higher cooling rates than could be achieved with the help of the above described spontaneous decay of individual particles. This is possible, when the time evolution of the system remains restricted onto a highly symmetric and strongly coupling subspace of states throughout the whole cooling process. Maximum cooling is obtained when the particles exhibit {\em cooperative} behavior in the excitation step as well as in the de-excitation step. To achieve this we assume as in Ref.~\cite{cool}, that the outward energy dissipation is conducted by an optical cavity. As shown in Fig.~\ref{setup}, the particles should be placed inside a resonant optical cavity with a relatively large rate for the leakage of photons through the cavity mirrors. The particles can then transfer their excitation collectively into the resonator field, from where the energy leaves the system without affecting the number of phonons in the setup and without changing the symmetry of the involved states. 

Collective behavior of the system requires furthermore that the Rabi frequency $\Omega_\nu$ of the laser field for the cooling of a vibrational mode with frequency $\nu$ and the cavity coupling constant $g$ are for all particles (practically) the same. Initially, the particles should all be prepared in their ground state (in the large $N$ limit fluctuations can be neglected). As shown below, the collective states of the assembly of $N$ particles then experience a very strong coupling to the laser field as well as to the cavity mode \cite{cool}. As a consequence, the number of phonons decreases exponentially with a with as large as $\sqrt{N}$ times the single-particle coupling constants, which can be much higher than previously predicted cooling rates in comparable setups \cite{vuletic0,vuletic1,horak0,horak,peter,jaksch,el,gardiner,black}. 

The proposed cooling scheme might be used to cool a large number of particles very efficiently. So it should be applicable to the preparation of Bose Einstein condensates. Currently, these experiments mainly use evaporative cooling \cite{evaporative} which systematically removes those atoms with a relatively high temperature from the trap. Consequently, only a small percentage of the initially trapped atoms is finally included in the condensate. If one could instead cool all the atoms efficiently, yet at the same time avoid the loss of particles, it should become easier to experiment with large condensates. Cooling is also crucial for ion trap quantum computing, where the achievable gate operation times can depend primarily on the efficiency of the cooling of a common vibrational mode \cite{cold}. First cavity-cooling experiments involving many particles and observing enhanced cooling rates have already been performed \cite{vuletic,Nagorny}. 

In this paper, we consider some algebraic features of the dynamics ruling  the collective cooling of many particles and further clarify some aspects of the underlying mechanisms, which have already been studied in Ref.~\cite{cool}. A remarkable feature is, for example, the presence of a relatively large and negative coherence $k_3$. Instead of solving the time evolution of the system explicitly, we avoid certain approximations made in Ref.~\cite{cool} by referring to the Heisenberg picture. The system considered in this paper and its collective dynamics may be a paradigmatic example for other applications of physical interest, like the evolution of a system undergoing a continuous phase transition, yet preserving some specific features during such an evolution. 

\section{The experimental setup}

The cavity coupling constant $g$ is for all particles the same, when the particles distribute themselves in the antinodes of the resonator field, as discussed in Ref.~\cite{peter,black} (see Fig.~\ref{setup}(a)). Alternatively, a ring resonator could be used, as proposed in Refs.~\cite{peter,Nagorny,jaksch}. In the following we consider a collection of $N$ two-level particles (atoms, ions or molecules) with ground states $|0 \rangle_i$ and excited states $|1 \rangle_i$ (see Fig.~\ref{setup}(b)). The setup should be operated in a parameter regime, where
\begin{eqnarray} \label{par}
\kappa \sim \sqrt{N} g \, , ~ {\textstyle {1 \over 2}} \sqrt{N} \eta \Omega \gg \Gamma ~~ {\rm with} ~~
\Omega \equiv \Big( \sum_\nu \Omega_\nu^2 \Big)^{1/2} .
\end{eqnarray}
Here $\kappa$ is the decay rate of a single photon in the cavity mode, $\Gamma$ is the spontaneous decay rate of a particle in the excited state and $\eta$ denotes the Lamb-Dicke parameter characterizing the steepness of the trap \cite{PLK}. As in Ref.~\cite{cool}, we discuss in this paper the two extreme cases, namely the cooling of common vibrational modes and the cooling of the individual phonon modes in the absence of common vibrational modes. 

\subsection{Cooling of common modes}

In the following, $b_\nu$ denotes the annihilation operator for a phonon in the common vibrational mode with frequency $\nu$, while $c$ is the annihilation operator for the cavity photons and $\sigma_i = |0 \rangle_{ii} \langle 1|$ is the lowering operator for particle $i$. Using the rotating wave approximation \cite{RWA} and going over to the interaction picture with respect to the interaction-free Hamiltonian, the time evolution of the system can be described by the Hamiltonian
\begin{eqnarray} \label{HI0}
H_{\rm comm} &=& \hbar \sum_{i,\nu} \Big( {\textstyle {1 \over 2}} \eta \Omega_\nu \, \sigma_i^\dagger b_\nu + g \, \sigma_i^\dagger c + {\rm H.c.}  \Big) \, .
\end{eqnarray}
One way to simplify this Hamiltonian is to introduce the effective
phonon annihilation operator
\begin{eqnarray} \label{not}
b \equiv \sum_\nu (\Omega_\nu/\Omega) \, b_\nu ~~ {\rm with} ~~
[b,b^\dagger]= 1 \, ,
\end{eqnarray}
which allows to write the Hamiltonian (\ref{HI0}) as
\begin{eqnarray} \label{HI}
H_{\rm comm} &=& \hbar \sum_i \Big( {\textstyle {1 \over 2}} \eta
\Omega \, \sigma_i^\dagger b + g \, \sigma_i^\dagger c + {\rm
H.c.}  \Big) ~.
\end{eqnarray}

\subsection{Cooling of individual particles}

We now consider the operator $b_{\nu,i}$ for the
annihilation of the phonons of mode $\nu$ of particle $i$. Again
we assume that the corresponding laser Rabi frequencies
$\Omega_\nu$ are for all particles the same. Then the Hamiltonian
of the system equals in the interaction picture and in the
rotating wave approximation 
\begin{eqnarray} \label{H0}
H_{\rm indi} &=& \hbar \sum_{i,\nu} \Big( {\textstyle {1 \over 2}}
\eta \Omega_\nu \, \sigma_i^\dagger b_{\nu,i} + g \,
\sigma_i^\dagger c + {\rm H.c.}  \Big) \, .
\end{eqnarray}
Proceeding as above, introducing the effective phonon annihilation operator
\begin{eqnarray}
b_i \equiv \sum_\nu (\Omega_\nu/\Omega) \, b_{\nu,i} ~~ {\rm with} ~~
[b_i,b_i^\dagger]= 1
\end{eqnarray}
and using Eq.~(\ref{par}), the Hamiltonian (\ref{H0}) becomes
\begin{eqnarray} \label{H}
H_{\rm ind} &=& \hbar \sum_i \Big( {\textstyle {1 \over 2}} \eta
\Omega \, \sigma_i^\dagger b_i + g \, \sigma_i^\dagger c + {\rm
H.c.}  \Big) \, .
\end{eqnarray}
Although this Hamiltonian has some similarities with the Hamiltonian (\ref{HI}), it describes a physically different situation. Instead of coupling to a set of common vibrational modes, each particle sees his own set of phonons. 

\subsection{Spontaneous emission}

Spontaneous emission is described in the following  by the
master equation \cite{barnett}
\begin{eqnarray} \label{rho}
{\dot{\rho}} &=& -  {{\rm i} \over \hbar} \, [ H_{\rm I}, \rho ]
+ \kappa \, \big(c \rho c^\dagger - {\textstyle{1 \over 2}}
c^\dagger c \rho - {\textstyle{1 \over 2}} \rho c^\dagger c \big)
\nonumber \\
&& + \Gamma \sum_i  \big(\sigma_i \rho \sigma_i^\dagger -
{\textstyle{1 \over 2}} \sigma_i^\dagger \sigma_i \rho -
{\textstyle{1 \over 2}} \rho \sigma_i^\dagger \sigma_i \big)
\end{eqnarray}
with $H_{\rm I}$ being $H_{\rm comm}$ or $H_{\rm ind}$, respectively, $\kappa$ being the decay rate of a photon in the cavity and $\Gamma$ being the decay rate of the particle excited state. From this equation one can easily see that the dissipation of cavity photons reduces the energy in the system without affecting the state of the particles. Cavity decay therefore does not disturb the collective behavior of the system. However, this does not apply to the emission of photons from the particles, which can negatively interfere with the collective cooling process.

\section{Bosonic subsystems} \label{boss}

In case of the cooling of common vibrational modes, the setup consists effectively of {\em two} different subsystems. One is the two-level particle system. As an effect of the emergence of the cooperative behavior, the collection of particles manifests itself as a {\em bosonic} system. We will see below that a further consequence of this is the transition to a strong coupling regime, which in turn implies a much shorter time scale for the system evolution. The other subsystem is the bosonic system of phonons and photons with a continuous {\it conversion} of phonons into photons. We will see that it is convenient to consider a boson mode which is a superposition of them. This is analogous to the field-atom polariton of Hopfield \cite{hopfield,hopfield2}.

\subsection{Bosonic behavior of the particles}

Suppose the time evolution of the particles is solely governed by the operators $\sigma^+ = \sum_i \sigma_i^+$, $\, \sigma^- = \sum_i \sigma_i^-$ and $\sigma_3 = {1 \over 2} \sum_i (|1
\rangle_{ii} \langle 1| - |0 \rangle_{ii} \langle 0|)$. The
$\sigma_i$ are the Pauli matrices. These operators obey the
$SU(2)$ commutation relations of a {\em fermion}-like $N$-body
system
\begin{eqnarray} \label{su2}
[\sigma_3, \sigma^\pm]= \pm \sigma^\pm ~~ {\rm and} ~~ [\sigma^-,
\sigma^+] = - 2 \sigma_3 \, .
\end{eqnarray}
Under this condition, the time evolution of the system with all
particles initially prepared in $|0 \rangle_i$ remains
restricted to a subspace of highly symmetric particle Dicke states
(for more details see \cite{cool,Dicke,Arecchi,Radcliffe}).
Since we assume a large number $N$ of particles, small
fluctuations in the initial state of the system can be neglected.
Let us denote by $l$ the number of particles excited by the laser
action into the upper state $|1 \rangle$. Then it can be shown
that for $N \gg l$ the particle system can be described by the 
collective operators \cite{cool}
\begin{eqnarray} \label{col}
S^+ \equiv {1 \over \sqrt{N}} \sum_i \sigma_i^+ \, , ~~ S^- \equiv
{1 \over \sqrt{N}} \sum_i \sigma_i^- \, ,  ~~ S_3 \equiv \sigma_3
\, , ~~
\end{eqnarray}
with $\sigma_3 = S^+S^-  - {1\over 2}N$. In the large $N$ limit 
($N \gg l$), the operators in  Eq. (\ref{col}) obey the relations
\begin{equation} \label{e2}
[ S_3,S^\pm] = \pm S^\pm \, , ~~~ [S^-, S^+] = 1 \, .
\end{equation}
This means, the su(2) algebra (\ref{su2}) written in terms of $S^\pm$ and $S_3$ contracts in the large-$N$ limit to the (projective) e(2) (or Heisenberg-Weyl) algebra (\ref{e2}) \cite{Wigner,Vitiello,SUV}. 

The physical implication of the familiar commutator relations
(\ref{e2})  is that the particles behave no longer like {\em
individual fermions}. Instead, the time evolution generates the
excitation of bosonic modes, namely collective dipole waves, with
$S^\pm$ denoting the creation and annihilation operators of the
associated quanta obeying the usual commutation relation
(\ref{e2}). As a consequence, the collection of the single
two-level particles manifests itself as a {\em bosonic} system. In other words, the ladder of equally-spaced Dicke states approximates to a weakly-excited harmonic oscillator.

Using the operators (\ref{col}), the Hamiltonian (\ref{HI}) for
the cooling of {\em common} vibrational modes can simply be
written as
\begin{eqnarray} \label{XXX}
H_{\rm comm} &=& \hbar \big( x \, S^+b  +  y \, S^+c + {\rm H.c.} \big) \, ,
\end{eqnarray}
where we have introduced the notation
\begin{eqnarray}
x \equiv {\textstyle {1 \over 2}} \sqrt{N} \eta \Omega ~~ {\rm and} ~~
y \equiv \sqrt{N} g \, .
\end{eqnarray}
From this one sees that the time evolution of the
system is mainly governed by the parameters $x$, $y$ and $\kappa$,
which scale as $\sqrt{N}$. We thus have, as a consequence of the
emergence of the particle collective behavior, the transition to
the strong coupling regime, ${\textstyle {1 \over 2}} \eta \Omega
\rightarrow {\textstyle {1 \over 2}} \sqrt{N} \eta \Omega$ and $g
\rightarrow \sqrt{N} g$. The evolution of the system 
happens no longer on the time scale given by the parameters $
{\textstyle {1 \over 2}} \eta \Omega$ and $g$ but on a much
shorter time scale defined by ${\textstyle {1 \over 2}} \sqrt{N}
\eta \Omega$ and $\sqrt{N} g$.

\subsection{Photon-phonon exchange} \label{aaa}

Another factor that contributes significantly to the described
cooling mechanism is the {\it interface} or {\it ambivalent} role
played by the particles with respect to the phonons and the
photons in the setup. The only difference between the
particle-phonon and particle-photon interaction is the difference of the
coupling constants, given by $\sqrt{N} {\textstyle {1 \over 2}}
\eta \Omega$ and $\sqrt{N} g$, respectively. In some sense, the
particles act as an engine transforming phonon energy into photon
energy. This last one is then dissipated outward through the
coupling with the cavity which allows energy leakage. However, in
the absence of the leakage of photons through the cavity, the
inverse transformation, photons to phonons, is also possible in
principle. In such a situation, interesting interference and
coherence effects arise, which we analyze below in detail. 

Indeed, a closer look at Eq.~(\ref{XXX}) reveals that the
Hamiltonian for the cooling of common vibrational modes can
alternatively be written as
\begin{eqnarray} \label{YYY}
H_{\rm comm} &=& \hbar z \, S^+a + {\rm H.c.}
\end{eqnarray}
with
\begin{eqnarray} \label{a}
z \equiv \sqrt{x^2 + y^2} \, , ~~ a \equiv {1 \over z} \,
(xb + yc) ~~ {\rm and} ~~ [a,a^\dagger]=1 \, . ~~
\end{eqnarray}
Instead of interacting with the phonons and photons separately,
the particles see the boson mode with annihilation operator $a$
and number operator
\begin{eqnarray} \label{deaf}
a^\dagger a = {1 \over z^2} \, \big[ x^2 \, b^\dagger b + y^2 \,
c^\dagger c + yx \, k_3 \big]
\end{eqnarray}
with
\begin{eqnarray}
k_3 \equiv b^\dagger c + b c^\dagger \, .
\end{eqnarray}
Physically, the creation of bosons corresponding to $a^\dagger $ does not only affect the number of phonons and the
number of photons in the system. Inevitably, it also creates a coherence between the $b$ and the $c$ subsystem. This coherence $k_3$ provides a ``symmetric" channel for the phonon-photon energy transformation. However, the leakage of energy outside the cavity perturbs such a symmetric phonon-photon balancing due to the $k_3$ action. The system reacts by subsequent adjustments, trying to recover its lost balance. Crucial to such a re-adjustment mechanism is the difference in the time scales. The time scale for reaching the quasi-stationary state is of the order $1/N$, while the time scale for spontaneous decay is, as we see in the next Section, of order $1/\sqrt{N}$. 

\section{Collective cooling of common vibrational modes} \label{comm}

The time evolution of the system turns out to be highly
non-linear since the second and higher order derivatives of the
physical observables are much larger than their first order
derivatives. The system reaches a quasi-stationary state on a time
scale of the order $1/N$. The formation of this local equilibrium
is solely governed by the effect of the Hamiltonian $H_{\rm
comm}$. Let us therefore first consider the situation, where we can neglect
spontaneous emission and $\kappa \approx 0$ and $\Gamma \approx 0$.

\subsection{Conservation laws in the absence of dissipation} \label{low}

In this case, the time evolution of the system, governed by the
Hamiltonian (\ref{YYY}), results in a redistribution of population
between the bosonic subsystem described  by $a^\dagger a$ and the
particles described by $S^+S^-$. To analyse this process we
introduce the operators
\begin{eqnarray}
L_1 &\equiv& {\textstyle {1 \over 2}} (S^+ a + S^- a^\dagger) \, , \nonumber \\
L_2 &\equiv& - {\textstyle {{\rm i} \over 2}} (S^+ a - S^- a^\dagger)
\, , \nonumber \\
L_3 &\equiv& {\textstyle {1 \over 2}} (S^+ S^- - a^\dagger a )
\end{eqnarray}
with the familiar $SU(2)$ commutators
\begin{eqnarray} \label{angi}
[L_i,L_j] = {\rm i} \epsilon_{ijk} \, L_k \, .
\end{eqnarray}
With this notation, the Hamiltonian (\ref{YYY}) becomes
\begin{eqnarray} \label{L1}
H_{\rm comm} = 2 \hbar z \, L_1
\end{eqnarray}
and this, formally, simply generates a rotation around the 1-axis.
In such an algebraic picture,  we can immediately conclude that
the angular momentum $L_1$ and the total angular momentum
\begin{eqnarray} \label{comp}
{\underline L}^2 &=& L_1^2 + L_2^2 + L_3^2 \nonumber \\
&=& {\textstyle {1 \over 2}} (S^+S^- + a^\dagger a) \, \big[ 1 +
{\textstyle {1 \over 2}} (S^+S^- + a^\dagger a) \big]
\end{eqnarray}
are conserved during the time evolution of the system under the
considered conditions. Especially, the conservation  of
${\underline L}^2$ implies the conservation of the total number of
bosons, as accounted for by the number operator $S^+S^- +
a^\dagger a$.

When the number of particles in the excited state $|1 \rangle_i$
remains small compared to $N$, the commutator relations (\ref{e2})
hold. In such a case, we can assume that the expectation value of
$S^+S^- = {1 \over N} \sum_i \sigma_i^+ \sigma_i^-$ remains small.
Neglecting $S^+S^-$ in Eq.~(\ref{comp}) implies
\begin{eqnarray} \label{cons0}
{\textstyle {{\rm d} \over {\rm d}t}} (a^\dagger a) = 0 \, .
\end{eqnarray}
Using the definition (\ref{a}), this conservation law translates into
\begin{eqnarray} \label{cons}
{\textstyle {{\rm d} \over {\rm d}t}} \big( \, y^2 \,
b^\dagger b + x^2 \, c^\dagger c - xy \, k_3 \, \big) = 0 \, ,
\end{eqnarray}
which coincides with Eq.~(16) in Ref.~\cite{cool} and describes
the continual balance between the rate of change of the
expectation value of $k_3$ and the rate of change of the total
number of phonons and photons in the system.

To show that Eq.~(\ref{cons}) does not violate the conservation of
the number of particles in the setup, we remark that the system
obeys a second conservation law. Considering again the Heisenberg
picture, we indeed find
\begin{eqnarray} \label{energy}
{\textstyle {{\rm d} \over {\rm d}t}} (b^\dagger b + c^\dagger c)
= - {\textstyle {{\rm d} \over {\rm d}t}} (S^+S^-) =  - 2 z L_2 \, .
\end{eqnarray}
In the presence of many particles, with most of them remaining in
their ground state, the expectation
value of $S^+S^-$ remains negligible. This implies ${\textstyle
{{\rm d} \over {\rm d}t}} (S^+S^-)=0$ and, consequently, also
$L_2=0$ \cite{note2}. Inserting this into Eq.~(\ref{energy}), we obtain the particle number 
conservation law
\begin{eqnarray} \label{cons2}
{\textstyle {{\rm d} \over {\rm d}t}} \big( \, b^\dagger b +
c^\dagger c \, \big) = 0 \, .
\end{eqnarray}
Despite being a coherence, the expectation value of $k_3$ acts
like a population. This allows the system to obey conservation of
the particle number $a^\dagger a$ as well as conservation of 
$b^\dagger b + c^\dagger c$ by balancing the
coherence $k_3$ accordingly. In the following, we study the
conversion of phonons into cavity photons and creation of a
non-zero coherence $k_3$ in more detail.

\subsection{Quasi-stationary states} \label{fast}

In the absence of spontaneous emission, the system reaches after a
very short time a stationary state with constant expectation
values for $b^\dagger b$, $c^\dagger c$ and $k_3$. To calculate
the corresponding values of the cavity photon number and coherence
$k_3$ as a function of the phonon number, we use again the
Heisenberg picture, and obtain the second order differential
equations
\begin{eqnarray} \label{sec}
{\textstyle {{\rm d}^2 \over {\rm d}t^2}} (b^\dagger b) &=& - 2 x^2 \,
b^\dagger b - xy \, k_3 \, , \nonumber \\
{\textstyle {{\rm d}^2 \over {\rm d}t^2}} (c^\dagger c) &=& - 2 y^2 \,
c^\dagger c - xy \, k_3 \, , \nonumber \\
{\textstyle {{\rm d}^2 \over {\rm d}t^2}} k_3 &=& - 2 xy \,
\big( b^\dagger b + c^\dagger c \big) - z^2 \, k_3 \, .
\end{eqnarray}
These equations imply that the first order derivatives of the
operators $b^\dagger b$, $c^\dagger c$ and $k_3$ change on a time
scale of the order $1/N$, which is, for large $N$, much faster
than the time scale on which the time evolution of $b^\dagger b $,
$c^\dagger c$ and $k_3$ takes place. Since we are only interested
in the time dependence of the phonon number $b^\dagger b$, we can
safely assume that the first order derivatives ${\textstyle {{\rm
d} \over {\rm d}t}} (b^\dagger b)$, ${\textstyle {{\rm d} \over
{\rm d}t}} (c^\dagger c)$ and ${\textstyle {{\rm d} \over {\rm
d}t}} k_3$ adapt adiabatically to the state of the system. Setting
the right hand sides of the differential equations (\ref{sec})
equal to zero, we find
\begin{eqnarray} \label{m}
 c^\dagger c = {x^2 \over y^2} \, b^\dagger b ~~ {\rm and} ~~
 k_3 = - {2x \over y} \, b^\dagger b \, .
\end{eqnarray}
Moreover, Eq.~(\ref{m}) implies, due to the conservation laws
(\ref{cons}) and (\ref{cons2}), that
\begin{eqnarray} \label{Wednesday}
{\textstyle {{\rm d} \over {\rm d}t}} (b^\dagger b) =
{\textstyle {{\rm d} \over {\rm d}t}} (c^\dagger c) = {\textstyle
{{\rm d} \over {\rm d}t}} k_3 = 0 \, .
\end{eqnarray}
This means that, after a short time of order $1/N$, the system
reaches a quasi-stationary state with a constant ratio between the
expectation value of the coherence $k_3$ and the number of photons
inside the cavity, respectively, compared to the total number of
phonons in the system. To a very good approximation, these ratios
remains constant throughout the whole cooling process.

\subsection{Cooling equations}

In the absence of spontaneous emission, at most a redistribution
of  phonon energy can occur in the system \cite{Bartana}.
Efficient cooling and the irreversible removal of energy from the
system requires dissipation. We therefore now consider the effect
of dissipation  in more detail. Using the master equation
(\ref{rho}) we find
\begin{eqnarray} \label{trop}
{\textstyle {{\rm d} \over {\rm d}t}} (b^\dagger b) &=& {\rm i} x \,
(S^+ b - S^- b^\dagger)  \, , \nonumber \\
{\textstyle {{\rm d} \over {\rm d}t}} (c^\dagger c) &=& {\rm i} y \,
(S^+ c - S^- c^\dagger) - \kappa \, c^\dagger c  \, , \nonumber \\
{\textstyle {{\rm d} \over {\rm d}t}} k_3 &=&  {\rm i} y \,
(S^+ b - S^- b^\dagger) + {\rm i} x \, (S^+ c - S^- c^\dagger) \nonumber \\
&& - {\textstyle {1 \over 2}} \kappa \, k_3 \, .
\end{eqnarray}
The leakage of photons through the cavity mirrors with decay  rate
$\kappa$ not only decreases the number of photons in the cavity
mode but also affects the size of $k_3$.

\begin{figure}
\begin{minipage}{\columnwidth}
\begin{center}
\resizebox{\columnwidth}{!}{\rotatebox{0}{\includegraphics{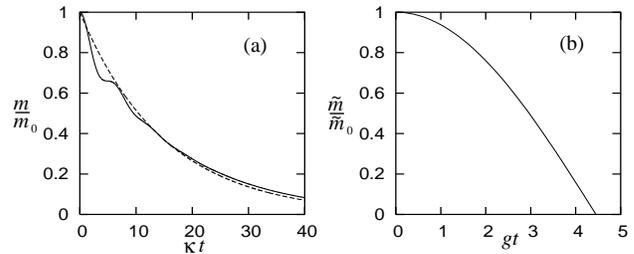}}}
\end{center}
\vspace*{-0.5cm} \caption{Cooling of common vibrational modes
obtained from a numerical solution of the master equation
(\ref{rho}) (solid line) in comparison to Eq.~(\ref{rate}) (dashed
line) for $g=10^{-3} \, \kappa$, $\eta\Omega=5 \cdot 10^{-4} \,
\kappa$, $N=10^6$ and $m_0=10^3$ (a). Cooling of individual phonon
modes, obtained from a numerical solution of Eq.~(21) in
Ref.~\cite{cool} under the assumption of negligible cavity photon
numbers, for $g=10^{-3} \, \kappa$, $\eta\Omega=5 \cdot 10^{-4} \,
\kappa$, $N=10^6$ and $\tilde m_0 = 10^9$ (b).} \label{num}
\end{minipage}
\end{figure}

In the previous subsection, we have seen that the system reaches a stationary state on a time scale of the order $1/N$ with $c^\dagger c$ and $k_3$ being a multiple of $b^\dagger b$. Using Eq.~(\ref{trop}), we find
\begin{eqnarray} \label{now}
{\textstyle {{\rm d} \over {\rm d}t}} (b^\dagger b) &=& - {x^2
\over y^2} \, \big[ \kappa \, c^\dagger c + {\textstyle {{\rm d} \over {\rm d}t}} \, (c^\dagger c) \big] +  {x \over y} \, \big[ {\textstyle {1 \over 2}} \kappa \, k_3 + {\textstyle {{\rm d} \over {\rm d}t}} k_3 \big] \, . \nonumber \\ &&
\end{eqnarray}
This shows that on the time scale we are interested in, namely a time scale of the order
$1/\sqrt{N}$, both, the operator $c^\dagger c$ and the presence of a negative valued operator $k_3$,  provide effective cooling channels in the system. Inserting the results (\ref{m}) and (\ref{Wednesday}) into Eq.~(\ref{now}) we obtain indeed
\begin{eqnarray} \label{ooo}
{\textstyle {{\rm d} \over {\rm d}t}} (b^\dagger b) &=& -
{x^2 z^2 \over y^4} \, \kappa \, b^\dagger b \, .
\end{eqnarray}
If $m$ denotes the phonon number expectation value $\langle
b^\dagger b \rangle_\rho$ and $m_0=m(0)$, we finally obtain
\begin{eqnarray} \label{rate}
m &=& m_0 \, \exp \Big[ - {x^2 z^2 \over y^4} \, \kappa t \Big] \, .
\end{eqnarray}
This describes exponential cooling of the atomic sample with  a
rate that can be as large as $\sqrt{N} g$ and ${1 \over 2}
\sqrt{N} \eta \Omega$ (cf.~Eq.~(\ref{par})). The result (\ref{rate}) coincides
with Eq.~(17) in Ref.~\cite{cool} and is in good agreement with an
exact numerical solution of the time evolution of the system (see
Fig.~\ref{num}(a)). Spontaneous emission from excited atomic
levels with decay rate $\Gamma$ is too slow to contribute to the cooling
of the system, since most of the particles remain in their ground
states and $\kappa \gg \Gamma$.

\subsection{Similarities between collective cooling and phase transitions}

Finally, we remark that the cooling process we describe here is
not simply a redistribution of phonons into cavity photons and
$k_3$, which then decay into the environment with the spontaneous
decay rates $\kappa$ and ${1 \over 2} \kappa$, respectively.
Instead, the leakage of photons through the cavity mirrors
disturbs the equilibrium expressed by Eqs.~(\ref{cons}) and
(\ref{cons2}) perturbing the otherwise conserved quantities
\begin{eqnarray} \label{wrong}
Q \equiv \langle a^\dagger a \rangle ~~ {\rm and} ~~
Q' \equiv \langle b^\dagger b \rangle + \langle c^\dagger c \rangle \, .
\end{eqnarray}
The coupling to the environment causes  a {\em dynamical response} in the system, which generates a transition from a state with a fixed ratio of phonons to cavity photons (cf.~Eq.~(\ref{m})) into a state with no phonons and photons in the setup.

In atom-cavity systems, the time evolution of the system is,  on a
very short time scale, usually dominated by spontaneous emission.
However, in the presence of many particles, the
evolution is primarily governed by the Hamiltonian $H_{\rm comm}$,
which drives the system into a quasi-stationary state
characterised by the preserved quantities $Q$ and $Q'$ within a
time of the order $1/N$. The presence of the spontaneous decay
rate $\kappa$, which scales as $\sqrt{N}$ (cf.~Eq.~(\ref{par})),
disturbs this stationary state and causes the system to continuously
assume new values for $Q$ and $Q'$. Using Eqs.~(\ref{m}) and
(\ref{ooo}), we find
\begin{eqnarray}
{\textstyle {{\rm d} \over {\rm d}t}} Q = - {x^2 z^2 \over y^4} \, \kappa \, Q ~~ {\rm and} ~~
{\textstyle {{\rm d} \over {\rm d}t}} Q' = - {x^2 z^2 \over y^4} \, \kappa \, Q' \, .
\end{eqnarray}
While the relative size of the photon number $\langle c^\dagger c
\rangle$ and the coherence $\langle k_3 \rangle$  with respect to
the number of phonons $m$ remain constant, their sum, as accounted
for by $Q$ and $Q'$ decrease exponentially in time due to the
presence of dissipation, namely leakage of photons through the
cavity mirrors. This is typical for a system undergoing a 
phase transition.

\section{Collective cooling of individual phonon modes} \label{ind}

There are many similarities but also many differences between the
cooling of the common and the cooling of the individual phonon
modes of particles. For example, one cannot simplify the
Hamiltonian $H_{\rm ind}$ using the collective lowering operator
$S$, as we did in Eq.~(\ref{XXX}). However, there is still a high
symmetry in the system and all particles are treated in exactly
the same way. To take this into account we introduce the vector
operators
\begin{eqnarray}
{\underline S^\pm} \equiv {\textstyle {1 \over \sqrt{N}}} \,
\big\{ \sigma_i^\pm \big\} \, , ~~
{\underline b} \equiv \big\{ b_i \big\} ~~ {\rm and} ~~ {\underline c}
\equiv \big\{ c \big\}
\end{eqnarray}
with the usual scalar product, such that, for example,
\begin{eqnarray}
{\underline S^\pm} \cdot {\underline b} = {1 \over \sqrt{N}} \,
\sum_i \sigma_i^\pm b_i \, .
\end{eqnarray}
This notation allows us to write the Hamiltonian (\ref{H}) as
\begin{eqnarray} \label{XXXX}
H_{\rm ind} &=& \hbar \, \big( x \, {\underline S}^+ \cdot
{\underline b} +  y \, {\underline S}^+ \cdot {\underline c} +
{\rm H.c.} \big) \, ,
\end{eqnarray}
which is of a similar form as the Hamiltonian $H_{\rm comm}$
in Eq.~(\ref{XXX}).

In analogy to Section \ref{aaa}, we proceed by introducing an
effective  annihilation operator $a_i$ for each particle $i$ with
\begin{eqnarray}
a_i \equiv {1 \over z} \, (xb_i + yc) \, , ~~  {\underline a}
\equiv \big\{ a_i \big\}
~~ {\rm with} ~~ [a_i, a_i^\dagger]=1 \, . ~~
\end{eqnarray}
Then the Hamiltonian (\ref{XXXX}) can be written as
\begin{eqnarray} \label{YYYY}
H_{\rm ind} &=& \hbar z \, \underline{S}^+ \cdot {\underline a}  +
{\rm H.c.}
\end{eqnarray}
Each particle couples individually to a new type of bosons, which
are a  superposition of a single phonon and a cavity photon. The
number operator accounting for all bosons $a_i$ equals
\begin{eqnarray}
{\underline a}^\dagger \cdot {\underline a} = {1 \over z^2} \,
\big[ x^2 \, {\underline b}^\dagger \cdot {\underline b} + y^2 \,
{\underline c}^\dagger \cdot {\underline c} + yx \, k_3 \big] \, ,
\end{eqnarray}
where we defined the coherence
\begin{eqnarray}
k_3 \equiv {\underline b}^\dagger \cdot {\underline c} + {\underline b}
\cdot {\underline c}^\dagger \, . ~~
\end{eqnarray}
While ${\underline b}^\dagger \cdot {\underline b}$ counts the
total  number of phonons in the system, the expectation value of
${\underline c}^\dagger \cdot {\underline c} = N c^\dagger c$.
Again, the coherence $k_3$ describes a certain ``symmetry''
between phonons and photons in the setup and accounts for a
continual conversion of the two types of bosons into each
other.

\subsection{Conservation laws in the absence of dissipation}

Neglecting spontaneous emission, i.e.~assuming $\kappa \approx 0$
and $\Gamma \approx 0$, and using the Hamiltonian (\ref{XXXX}), we
find
\begin{eqnarray} \label{trop2}
{\textstyle {{\rm d} \over {\rm d}t}} ({\underline b}^\dagger
\cdot {\underline b}) &=& {\rm i} x \, ({\underline S}^+ \cdot
{\underline b} - {\underline S}^- \cdot {\underline b}^\dagger)
\, , \nonumber \\
{\textstyle {{\rm d} \over {\rm d}t}} ({\underline c}^\dagger
\cdot {\underline c}) &=& {\rm i} N y \, ({\underline S}^+ \cdot
{\underline c} - {\underline S}^- \cdot {\underline c}^\dagger) \, , \nonumber \\
{\textstyle {{\rm d} \over {\rm d}t}} k_3 &=&  {\rm i} y \,
({\underline S}^+ \cdot {\underline b} - {\underline S}^-
\cdot {\underline b}^\dagger) + {\rm i} x \, ({\underline S}^+
\cdot {\underline c} - {\underline S}^- \cdot {\underline c}^\dagger) \, .
\nonumber \\ &&
\end{eqnarray}
In the derivation of these equations, we neglected the operators $\sigma_i^\dagger b_j$ and
$\sigma_i b_j^\dagger$  with $i \neq j$ since there are no interactions between particles
and the phonons of other particles. From Eq.~(\ref{trop2}) we see
that there are, as before, two conserved quantities in the system,
namely
\begin{eqnarray}
Q &\equiv & y^2 \, \langle {\underline b}^\dagger \cdot {\underline b}
\rangle + {x^2 \over N} \, \langle {\underline c}^\dagger \cdot
{\underline c} \rangle  - xy \, \langle k_3 \rangle \, , \nonumber \\
Q' &\equiv & \langle {\underline b}^\dagger \cdot {\underline b}
\rangle + {1 \over N} \, \langle {\underline c}^\dagger \cdot
{\underline c} \rangle  \, .
\end{eqnarray}
One is associated with the total number of bosonic particles  with
annihilation operators $a_i$, the other one counts the total
number of phonons and photons in the setup and
\begin{eqnarray} \label{lee}
{\textstyle {{\rm d} \over {\rm d}t}} Q = {\textstyle {{\rm d}
\over {\rm d}t}} Q' = 0 \, .
\end{eqnarray}
From the formal equivalence of the Hamiltonians (\ref{YYY})  and
(\ref{YYYY}) pointed out in the previous subsection, one might
have had expected that the expectation value $\langle {\underline
a}^\dagger \cdot {\underline a} \rangle$ is preserved in the time
evolution of the system. However, this would only be the case if
$a_i$ and $a_j^\dagger$ commute with each other for $i \neq j$
and does not apply here.

\subsection{Quasi-stationary states}

To calculate the distribution of phonons, cavity photons and
coherence $k_3$ in the system, which builds up in the absence of
spontaneous emission, we now consider the second derivatives of
the operators ${\underline b}^\dagger \cdot {\underline b}$,
~${\underline c}^\dagger \cdot {\underline c}$ and $k_3$ in the
Heisenberg picture. Neglecting again the population in the excited
states, as accounted for by ${\underline S}^+ \cdot {\underline
S}^-$, we obtain
\begin{eqnarray} \label{sec2}
{\textstyle {{\rm d}^2 \over {\rm d}t^2}} ({\underline b}^\dagger
\cdot {\underline b})
&=& - {2 x^2 \over N} \, {\underline b}^\dagger  \cdot {\underline b} -
{xy \over N} \, k_3 \, , \nonumber \\
{\textstyle {{\rm d}^2 \over {\rm d}t^2}} ({\underline c}^\dagger \cdot
{\underline c})
&=& - 2 y^2 \, {\underline c}^\dagger \cdot {\underline c} - xy \, k_3 \, ,
\nonumber \\
{\textstyle {{\rm d}^2 \over {\rm d}t^2}} k_3
&=& - {2 xy \over N} \, \big( {\underline b}^\dagger  \cdot
{\underline b} + {\underline c}^\dagger  \cdot {\underline c} \big) -
{z^2 \over N} \, k_3 \, . ~~~
\end{eqnarray}
Setting the right hand side of these equations equal  to zero and
taking the conservation laws (\ref{lee}) in the absence of
dissipation into account, we find that the system
possesses a stationary state with
\begin{eqnarray} \label{m2}
&& {\underline c}^\dagger \cdot {\underline c} = {x^2
\over y^2} \, {\underline b}^\dagger \cdot {\underline b} \, , ~~ k_3 = - {2x \over y} \, {\underline b}^\dagger \cdot {\underline b} \, , \nonumber \\
&& {\textstyle {{\rm d} \over {\rm d}t}} (b^\dagger b) =
{\textstyle {{\rm d} \over {\rm d}t}} (c^\dagger c) =
{\textstyle {{\rm d} \over {\rm d}t}} k_3 = 0 \, .
\end{eqnarray}
In contrast to the results in Section \ref{fast}, this  equation
describes a state with only a relatively small number of photons
in the cavity mode compared to the total number of phonons in the
setup. However, as we see in the next subsection, the presence of a
negative coherence $k_3$ of the same order of magnitude as
${\underline b}^\dagger \cdot {\underline b}$ provides
a cooling channel, which is sufficient to obtain
a cooling rate of the same order of magnitude as the cavity decay
rate $\kappa$.

\subsection{Cooling equations}

As before, the effective removal of phonons from the
setup requires leakage of photons through the cavity mirrors. To
take spontaneous emission into account we consider again the
master equation (\ref{rho}) and find
\begin{eqnarray} \label{trop3}
{\textstyle {{\rm d} \over {\rm d}t}} ({\underline b}^\dagger \cdot
{\underline b})
&=& {\rm i} x \, ({\underline S}^+ \cdot {\underline b} - {\underline S}^-
\cdot {\underline b}^\dagger)  \, , \nonumber \\
{\textstyle {{\rm d} \over {\rm d}t}} ({\underline c}^\dagger \cdot {\underline c})
&=& {\rm i} N y \, ({\underline S}^+ \cdot {\underline c} - {\underline S}^-
\cdot {\underline c}^\dagger) - \kappa \, {\underline c}^\dagger \cdot
{\underline c}  \, , \nonumber \\
{\textstyle {{\rm d} \over {\rm d}t}} k_3 &=&  {\rm i} y \, ({\underline S}^+
\cdot {\underline b} - {\underline S}^- \cdot {\underline b}^\dagger) + {\rm i} x \,
({\underline S}^+ \cdot {\underline c} - {\underline S}^- \cdot {\underline c}^\dagger)
\nonumber \\
&& - {\textstyle {1 \over 2}} \kappa \, k_3 \, ,
\end{eqnarray}
which implies
\begin{eqnarray} \label{now2}
{\textstyle {{\rm d} \over {\rm d}t}} ({\underline b}^\dagger \cdot {\underline b})
&=& - {x^2 \over Ny^2} \, \big[ \kappa \, {\underline c}^\dagger \cdot
{\underline c} + {\textstyle {{\rm d} \over {\rm d}t}} \,
({\underline c}^\dagger \cdot {\underline c}) \big]  \nonumber \\
&& +  {x \over y} \, \big[ {\textstyle {1 \over 2}} \kappa \,
k_3 + {\textstyle {{\rm d} \over {\rm d}t}} k_3 \big] \, .
\end{eqnarray}
As already mentioned above, the number of photons in the cavity
remains  relatively small and does not contribute considerably to
the cooling process. Neglecting ${\underline c}^\dagger \cdot
{\underline c}$ and using the results from the previous
subsection, we obtain
\begin{eqnarray} \label{now2}
{\textstyle {{\rm d} \over {\rm d}t}} ({\underline b}^\dagger \cdot {\underline b})
&=& - {x^2 \over y^2} \, \kappa \, {\underline b}^\dagger \cdot {\underline b}
\end{eqnarray}
and
\begin{eqnarray} \label{rate2}
\tilde m &=& \tilde m_0 \, \exp \Big[ - {x^2 \over y^2} \, \kappa t \Big] \, ,
\end{eqnarray}
where $\tilde m$ and $\tilde m_0$ now describe the total number
of phonons in the system, as accounted for by ${\underline
b}^\dagger \cdot {\underline b}$ at time $t$ and $t=0$,
respectively. As in the case of the cooling of common vibrational
modes, cooling rates of the same order of magnitude as the cavity
decay rate $\kappa$ can be obtained, which can be as large as
$\sqrt{N}$ times the single particle coupling constants.

However, there are some differences with respect to the case
considered in Section \ref{comm}. In the present case,
achieving such a high cooling rate and cooling the system to very
low temperatures first requires the build up of a reasonably large
coherence $k_3$, which does not exist in the absence of any laser
driving. Suppose, there are no initial correlations between the
particles and their motional degrees of freedom. Then ${\textstyle
{{\rm d} \over {\rm d}t}} k_3 =0$ (cf.~Eq.~(\ref{trop3})).
Moreover, the second order derivative of $k_3$ increases only on a
time scale of order one (cf.~Eq.~(\ref{sec2})). Initially, the
system is far away from the quasi-stationary state described in
the previous subsection. However, once $k_3$ reaches its
equilibrium, the collective cooling process can begin. The
coherence $k_3$ needs no longer to be established; the system only
has to adapt to the small changes of $Q$ and $Q'$ caused by the
leakage of photons through the cavity mirrors. For a numerical
solution of the time evolution of the system see Fig.~\ref{num}(b).

\section{Conclusions}

The emergence of collective dynamics in a system of a large number
of particles manifests itself in some macroscopic features in the
system behavior. As an example, we considered the problem of fast and
efficient cooling of an assembly of $N$ two-level particles
trapped inside a leaky optical cavity. Results obtained here confirm
those derived in Ref.~\cite{cool}. The particles are excited by 
red-detuned laser fields. When the coupling constants are for all
particles the same, a collective behavior emerges and the cooling
rate can be as large as $\sqrt{N}$ times the single-particle
coupling constants. The generation of cooperative behavior of the
$N$ particles is crucial in the excitation step as well as in the
de-excitation step. The collective states  of the assembly of $N$
particles then experience a very strong coupling to the laser field as
well as to the cavity mode. 

We have considered the case of particles sharing a common phonon 
mode and the case of individual phonon modes for each particle. The 
two cases are similar and in both cases the collective cooling rate is very 
fast as compared to the single-particle cooling rate. The two cases are, however,
different for the behavior of the coherence $k_3$: in the
individual mode case, high cooling rates are achieved by first
building up a reasonably large coherence $k_3$, which does not
exist in the absence of any laser driving. Initially, the system
is far away from the quasi-stationary state, where the phonon and
the photon populations are balanced through the $k_3$ action. Only
when $k_3$ reaches its equilibrium, the collective cooling process
can begin and the system can adapt to the small changes of $Q$
and $Q'$ caused by the leakage of photons through the cavity
mirrors. In the common mode case, there is no need for the
``pre-cooling" phase for a build up of the coherence $k_3$.

The system considered in this paper  may be a paradigmatic example
for other  applications of physical interest, such as systems
undergoing continuous phase transitions, yet preserving some
specific features during their evolution. Systems presenting such
a behavior might be of interest as well in biology.

\section{Acknowledgments}

This work was supported in part by the European   Union, COSLAB
(ESF Program), INFN, INFM  and the UK Engineering and Physical
Sciences Research Council.  A.B. acknowledges funding as a James
Ellis University Research Fellow from the Royal Society and the
GCHQ.


\begin{thebibliography}{00}
\bibitem{RWA}
This approximation introduces some errors when calculating  the
behavior of the system for times $t \ll 1/ \nu$, with a
restriction of achievable cooling rates from above. Nevertheless,
collective cooling can be much more efficient than previously
considered mechanisms.
\bibitem{note2}
Assuming $L_2 = -{{\rm i} \over 2} (S^+a - S^- a^\dagger) =0$  is
also equivalent to neglecting the coherence operators $k_1 \equiv
S^+b - S^- b^\dagger$ and $k_2 \equiv S^+c - S^- c^\dagger$, which
is consistent with our calculations in Ref.~\cite{cool}. There we
showed that $k_1$ and $k_2$ remain indeed zero to a very good
approximation.
\end{thebibliography}

\end{document}